\documentclass[paper]{JHEP3}

\title{Gauge/string duality and scalar glueball mass ratios}

\author{Henrique Boschi-Filho\\
Instituto de F\'{\i}sica, Universidade Federal do Rio de Janeiro, 
Caixa Postal 68528, RJ 21945-970 -- Brazil\\ 
E-mail: \email{boschi@if.ufrj.br}}

\author{Nelson R. F. Braga\\
Instituto de F\'{\i}sica, Universidade Federal do Rio de Janeiro, 
Caixa Postal 68528, RJ 21945-970 -- Brazil\\
E-mail: \email{braga@if.ufrj.br}}
 
\abstract{It has been shown by Polchinski and Strassler that 
the scaling of high energy QCD scattering amplitudes can be obtained 
from  string theory. They considered an AdS slice as an 
approximation for the dual space of a confining gauge theory. 
Here we use this approximation to estimate in a very simple way 
the ratios of scalar glueball masses imposing Dirichlet boundary 
conditions on the string dilaton field. 
These ratios are in good agreement with the results in the literature.
We also find that they do not depend on the size of the slice.}

\keywords{fth, ads}

\preprint{hep-th/0212207}

\begin{document}


Glueballs are bound states of gluons as predicted by QCD,  
presently the main candidate theory to describe the strong 
interactions.
These bound states have not yet been observed but hopefully 
will appear in near future accelerators. The investigation of such 
states using QCD in the low energy (strong coupling) regime, 
in particular looking for their masses,  
requires a non perturbative approach.
In this case lattice calculations produce 
interesting results (see for instance\cite{LAT1,LAT2} 
and references therein).

An alternative approach to strong interactions is based on the
idea that they have a description in terms of strings\cite{Planar,Pol}. 
A remarkable step in this direction was given by Maldacena\cite{Malda} 
proposing the equivalence between conformal fields  and string
theory in anti-de Sitter spacetime (AdS/CFT correspondence)
\cite{GKP,Wi,Malda2}.
In particular glueball operators of the conformal gauge theory defined on 
the AdS boundary are in correspondence with the string dilaton field.
  
The description of strong interactions based on this correspondence  
requires the breaking of conformal invariance, which can be done 
in different ways.  
Witten \cite{Wi2} proposed a formulation of the correspondence in such 
a way that the AdS space accommodates a  Schwarzschild black hole.
This procedure introduces a scale breaking conformal invariance. 
Then it is possible to obtain the ratio of glueball masses from 
supergravity approximation to string theory.
The corresponding supergravity equations do not allow analytic 
solutions but the eigenvalues related to the glueball masses 
can be found using a WKB method \cite{MASSG}.

Another possibility to break conformal invariance is to consider 
a slice (actually two slices sticked together) of the AdS space 
as in the Randall-Sundrum model\cite{RS1,RS2} which proposes a 
unification of the weak and Planck scales.
Such an AdS slice was used recently by Polchinski and Strassler
\cite{PS} as an approximation for the string dual space corresponding to 
strong interactions (see also\cite{GI,BB3,BT,AN}).
They used this model to reproduce the scaling  of high energy glueball 
scattering amplitudes as predicted by QCD\cite{QCD1,BRO}. 
Following the Polchinski and Strassler proposal we consider an AdS slice 
assuming that there is still a bulk/boundary correspondence, 
in particular between scalar glueballs and the dilaton.
Imposing boundary conditions the string dilaton field acquires 
discrete modes. 
Then it is natural to relate the spectrum of the dilaton field to the 
masses  of the scalar glueballs. Our results are in good agreement with
lattice and supergravity calculations (see tables 1 and 2).

According to the AdS/CFT correspondence string theory defined on 
AdS$_5 $  times a transverse space $X$ is dual to the large $N$ limit 
of  $SU(N)$ conformal gauge theories with extended supersymmetry defined on 
the four dimensional boundary. The metric for this space can be written as
\begin{equation}
\label{metric}
ds^2=\frac {R^2 }{ z^2}\Big( dz^2 \,+(d\vec x)^2\,
- dt^2 \Big) \,+ \,R^2 ds^2_X \,\,,
 \end{equation}
 
\noindent where $z, \vec x , t\,$ are the Poincare coordinates that 
describe the AdS space with radius $R$ and $ds^2_X$ corresponds to 
the metric of a convenient transverse space, as for example $S^5$.  

We consider an AdS slice corresponding to the region
$0 \le z \le z_{max}\,$.
According to the AdS/CFT duality, there is a holographic relation 
between bulk and boundary theories such that low energies 
correspond to large $z$-values. Then $z_{max} $ is an infrared cut 
off for the nonconformal boundary theory. 
Further we take the dilaton momenta associated with the directions of 
the space $X$ to be negligible.
Then choosing Dirichlet boundary condition the free dilaton field at 
$ z = z_{max}$ can be cast in terms of the Bessel function $J_2$ 
into the form \cite{BB1}
\begin{eqnarray}
\label{QF}
\Phi(z,\vec x,t) &=& \sum_{p=1}^\infty \,
\int { d^3 k \over (2\pi)^{3}}\,
{z^{2} \,J_2 (u_p z ) \over z_{max}\,\, w_p(\vec k ) 
\,J_{3} (u_p z_{max} ) }\nonumber\\
&\times& \lbrace { {\bf a}_p(\vec k )\ }
 e^{-iw_p(\vec k ) t +i\vec k \cdot \vec x}\,
\,+\,\,h.c.\rbrace\,.
\end{eqnarray}

\noindent 
Here $w_p(\vec k ) \,=\,\sqrt{ u_p^2\,+\,{\vec k}^2}\,$, $h.c.\,$ 
means hermitian conjugate and  $u_p$ are the discrete momenta  
associated with coordinate $z$,  defined by 
\begin{equation}
\label{up}
u_p \,=\,\frac{ \chi_{_{2\,,\,p}}}{z_{max}}\;,
\end{equation}

\noindent where $ J_2 (\chi_{_{2\,,\,p}} )=0\,$, and the operators 
${\bf a}_p\, ,\;{\bf a}^{\dagger}_p \,$ satisfy canonical 
commutation relations.
 
 On the boundary ($ z = 0)$ of the AdS slice we consider composite 
operators representing glueballs with masses $\mu_p$. 
As in ref. \cite{PS} we associate the size $z_{max}$ of the AdS slice 
with the mass of the lightest glueball  $\mu_1$  
\begin{equation}
\label{up2}
z_{max} \,=\, { C \over \mu_1}\,,
\end{equation}

\noindent where $C$ is an arbitrary constant related to the choice of  
$z_{max}$. From equations (\ref{up}) and (\ref{up2}) we have 
\begin{equation}
\label{up3}
u_p \,=\,{\chi_{2\,,\,p}\over C} \,\, \mu_1 .
\end{equation}

Then the gauge/string correspondence suggests that the glueball masses
are proportional to the discrete dilaton modes: 

\begin{equation}
{u_p \over \mu_p }\,=\,C^\prime\,\,, 
\end{equation}
 
\noindent 
where $C^\prime \,=\, \chi_{2\,,\,1}/ C\,$ according to eq. (\ref{up3}).

So the glueball masses are related to the zeros of the Bessel
functions by
\begin{equation}
\label{QCD4}
{ \mu_p\over \mu_1 }\,=\,{\chi_{2\,,\,p}\over \chi_{2\,,\,1}}\,\,.
\end{equation}

\noindent This is our main result. It is remarkable that it is
independent of the size $z_{max}$, although the individual masses
depend on this cut off. 
 
Using the values of the zeros of the Bessel function, 
we find the mass ratios for the scalar glueball state 
$J^{PC}\,=\,0^{++}$ and its excitations. 
Our results are presented in table 1 together with lattice
\cite{LAT1,LAT2} and AdS-Schwarzschild black hole supergravity
\cite{MASSG} calculations. 
From this table one can see that our simple approximation is in good 
agreement with these previous calculations.

\TABLE{
\centering
\begin{tabular}{l|ccc}
4d Glueball  & lattice, $N=3$ &
AdS-BH & AdS slice \\
 \hline
 $0^{++}$ & $1.61 \pm 0.15$   & 1.61 {\rm (input)} & 1.61 {\rm (input)} \\
 $0^{++*}$ &  2.8   & 2.38 & 2.64 \\
 $0^{++**}$ &   - & 3.11 & 3.64 \\
 $0^{++***}$ &  -  & 3.82 & 4.64\\
 $0^{++****}$ &  -  & 4.52 & 5.63\\
 $0^{++*****}$ &  -  & 5.21 & 6.62\\
\end{tabular}
\parbox{6in}{\caption{ 
\sl Masses of the first few $0^{++}$ four dimensional 
glueballs with $SU(N)$ and $N=3$, in GeV, from lattice QCD\cite{LAT1,LAT2}, 
from AdS-Schwarzschild black hole supergravity (AdS-BH)\cite{MASSG} 
and our results from AdS slice normal modes eq. (\ref{QCD4}).}}}

\bigskip

A similar approach can be used to  estimate the glueball masses in 
QCD$_3$. In this case we consider  AdS$_{4}$ and the dilaton fields 
are expanded in terms of the Bessel function $J_{3/2} $ and the 
mass ratios for the $3$ dimensional "glueballs"  are given by 
\begin{equation}
\label{QCD3}
{ \mu_p\over \mu_1 }\,=\,{\chi_{3/2\,,\,p}\over \chi_{3/2\,,\,1}}\,\,.
\end{equation}
\noindent 
Using this relation we obtain the ratio of masses presented in table 2 
together with lattice and $AdS$-Schwarzschild black hole supergravity 
calculations. The agreement here is also good.

\bigskip

\TABLE{
\centering
\begin{tabular}{l|cccc}
3d Glueball & lattice, $N=3$ & lattice, $N\rightarrow \infty$ &
 AdS-BH& AdS slice \\
 \hline
 $0^{++}$ & $4.329 \pm 0.041$ & $4.065 \pm 0.055$ & 4.07 ({\rm input})
& 4.07 ({\rm input}) \\
 $0^{++*}$ & $6.52 \pm 0.09$ & $6.18 \pm 0.13$ & 7.02 & 7.00\\
 $0^{++**}$ & $8.23 \pm 0.17$ & $7.99 \pm 0.22$ & 9.92 & 9.88 \\
 $0^{++***}$ &  - & - & 12.80 & 12.74 \\
 $0^{++****}$ &  - & - & 15.67 & 15.60\\
 $0^{++*****}$ & -  & - & 18.54 & 18.45\\
\end{tabular}
\parbox{6in}{\caption{
\sl $0^{++}$ three dimensional glueball masses with 
$SU(N)\,$ from lattice QCD\cite{LAT1,LAT2} (in units of string tension), 
from AdS-Schwarzschild black hole supergravity (AdS-BH)\cite{MASSG} 
and our results from AdS slice normal modes, eq. (\ref{QCD3}). }}}

The relations (\ref{QCD4},\ref{QCD3}) for the ratio of glueball masses 
were found by us before in \cite{BB4} using a mapping between
bulk and boundary quantum states.   
Here we have seen that these results do not depend on any particular  
mapping between quantum states.

In conclusion we have seen that string theory defined in an AdS slice 
can be applied to estimate the scalar glueball mass ratios in a very
simple way. We hope that such an approach can be used to  
estimate the mass ratios of other strongly interacting states.

\acknowledgments{
We would like to thank J. R. T. Mello Neto for important discussions.
The authors are partially supported by CNPq, FINEP , CAPES (PROCAD) 
and FAPERJ - Brazilian research agencies.}



\end{document}